\documentclass[a4paper]{article}
\usepackage{amsmath}
\usepackage{graphicx}
\usepackage{cite}
\usepackage{hyperref}

\begin{document}
\title{Iterated maps for clarinet-like systems}
\author{P.-A. Taillard \\
 Conservatoire de Musique Neuch\^{a}telois, Avenue
L\'{e}opold-Robert 34,\\ 2300 La Chaux-de-Fonds, Switzerland, \\
J. Kergomard\footnote{Tel 33 491164381, Fax 33 491228248,
kergomard@lma.cnrs-mrs.fr}
\\ Laboratoire de M\'{e}canique et d'Acoustique, CNRS UPR 7051, 31
Chemin\\ Joseph Aiguier, 13402 Marseille Cedex 20, France \\ and
 F. Lalo\"{e} \\
 Laboratoire Kastler-Brossel, ENS,\\ UMR 8552 CNRS, ENS, et UMPC; 24 rue Lhomond, 75005 Paris, France }

 \maketitle
\begin{abstract}
The dynamical equations of clarinet-like systems are known to be
reducible to a non-linear iterated map within reasonable
approximations. This leads to time oscillations that are
represented by square signals, analogous to the Raman regime for
string instruments. In this article, we study in more detail the
properties of the corresponding non-linear iterations, with
emphasis on the geometrical constructions that can be used to
classify the various solutions (for instance with or without reed
beating) as well as on the periodicity windows that occur within
the chaotic region. In particular, we find a regime where period
tripling occurs and examine the conditions for intermittency. We
also show that, while the direct observation of the iteration
function does not reveal much on the oscillation regime of the
instrument, the graph of the high order iterates directly gives
visible information on the oscillation regime (characterization of
the number of period doubligs, chaotic behaviour, etc.).
\end{abstract}


\section{Introduction}

Non-linear iterated maps are now known as an universal tool in numerous
scientific domains, including for instance mechanics, hydrodynamics and
economy \cite{May} \cite{Berge-Pomeau-Vidal} \cite{Collet-Eckmann}.\ They
often appear because the differential equations describing the dynamics of a
system can be reduced to non-linear iterations, with the help of
Poincar\'{e} recurrence maps for instance.\ The resulting iterations combine
a great mathematical simplicity, which makes them convenient for numerical
simulations, with a large variety of interesting behaviors, providing
generic information on the properties of the system.\ In particular, they
are essential to characterize one of the routes to chaos, the cascade of
period doublings \cite{Feigenbaum:79}.

In musical acoustics, Mc Intyre \textit{et al.} have given, in a
celebrated article \cite{McIntyre:83}, a general frame for
calculating the oscillations of musical instruments, based upon
the coupling of a linear resonator and a non-linear excitator (for
reed instruments, the flow generated by a supply pressure in the
mouth and modulated by a reed). In an appendix of their article
they show that, within simplified models of self-sustained
instruments, the equations of evolution can also be reduced to an
iterated map with appropriate non-linear functions. For resonators
with a simple shape such as a uniform string or a cylindrical
tube, the basic idea is to choose variables that are amplitudes of
the incoming and outgoing waves (travelling waves), instead of
usual acoustic pressure and volume velocity in the case of reed
instruments. If the inertia of the reed is ignored (a good
approximation in many cases), and if the losses in the resonator
are independent of frequency, the model leads to simple
iterations; the normal oscillations correspond to the so called
\textquotedblleft Helmholtz motion\textquotedblright , a regime in
which the various physical quantities vary in time by steps, as in
square signals. Square signals obviously are a poor approximation
of actual musical signals, but this approach is sufficient when
the main purpose is to study regimes of oscillation, not
tone-color.

 In the case of clarinet-like systems, the idea was then expanded
\cite {Maganza:86}, giving rise to experimental observations of
period doubling scenarios and to considerations on the relations
between stability of the regimes and the properties of the second
iterate of the non-linear function; see also \cite{Brod:90} and
especially \cite{Kergomard:95} for a review of the properties of
iterations in clarinet-like systems and a discussion of the
various regimes (see also \cite{lizee}). More recent work includes
the study of oscillation regimes obtained in experiments
\cite{Idogawa:93, Gibiat:00}, computer simulation
\cite{Kergomard:04} as well as theory \cite{Ollivier:05,
Dalmont:05}.

The general form of the iteration function that is relevant for
reed musical instruments is presented in section \ref{iteration}.
It it is significantly different from the usual iteration parabola
(i.e. the so-called logistic map).\ Moreover, it will be discussed
in more detail that the control parameters act in a rather
specific way, translating the curve along an axis at $45^{\circ }$
rather than acting as an adjustable gain.

The purpose of the present article is to study the iterative
properties of functions having this type of behavior, and their
effect on the oscillation regimes of reed musical instruments.\ We
will study the specificities and the role of the higher order
iterates of this class of functions, in particular in the regions
of the so called ``periodicity windows'', which take place beyond
the threshold of chaos.\ These windows are known to contain
interesting phenomena \cite{Berge-Pomeau-Vidal, Vallee, Stef:99},
for instance period tripling or a route to intermittence, which to
our knowledge have not yet been studied in the context of reed
musical instruments.\ Moreover, the iterates give a direct
representation of the zones of stability of the different regimes
(period doublings for instance), directly visible on the slope of
the corresponding iterate.

For numerical calculations, it is necessary to select a particular
representation of the non-linear function, which in turn requires to choose
a mathematical expression of the function giving the volume flow rate as a
function of the pressure difference across the reed.\ A simple and realistic
model of the quasi-static flow rate entering a clarinet mouthpiece was
proposed in 1974 by Wilson and Beavers \cite{Wilson}, and discussed in more
detail in 1990 by Hirschberg \textit{et al.} \cite{Hirschberg:90}. This
model provides a good agreement with experiments \cite{Dalmont:03} and leads
to realistic predictions concerning the oscillations of a clarinet \cite
{Dalmont:07}. Using this mathematical representation of the flow rate, we
will see that iterations lead to a variety of interesting phenomena.\ Our
purpose here is not to propose the most elaborate possible model of the
clarinet, including all physical effects that may occur in real
instruments.\ It is rather to present general ideas and mathematical
solutions as illustration of the various class of phenomena that can take
place, within the simplest possible formalism; in a second step, one can
always take this simple model as a starting point, to which perturbative
corrections are subsequently added in order to include more specific details.

We first introduce the model in \S\ \ref{model}, and then discuss
the properties of the iteration function in \S\ \ref{properties}.\
The bifurcations curves are obtained in \S\ \ref{bifurcations}
and, in \S\ \ref {iterated}, we discuss the iterated functions and
their applications in terms of period tripling and intermittence.
In particular we see how the graph of high order iterates give
visible information on the regime of oscillation (number of period
doublings for instance) or the appearance of a chaotic regime,
while nothing special appears directly  in the graph of the first
iterate.\ Two appendices are added at the end.

\section{The model}

\label{model}We briefly recall the basic elements of the model, the
non-linear characteristics of the excitator, and the origin of the
iterations within a simplified treatment of the resonator.

\subsection{Nonlinear characteristics of the entering flow}

\begin{figure}[h]
\centering \includegraphics[width=10cm]{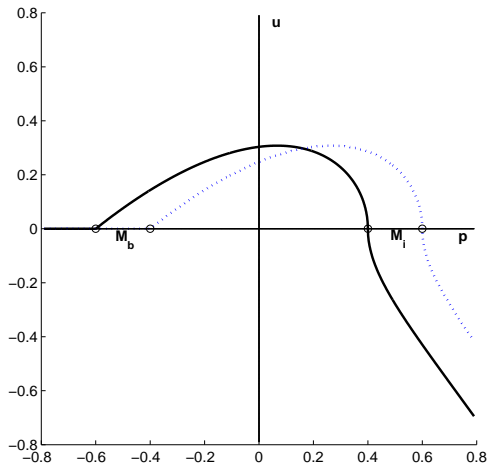} \caption{Graph
showing the air flow rate $u$ entering the resonator of a clarinet
as a function of the internal pressure $p$ (pressure in the
mouthpiece). All physical quantities are expressed in
dimensionless units, as explained in the text. M$_b$ corresponds
to the contact point where the internal pressure in the mouthpiece
bends the reed sufficiently to close the
channel, so that the flow vanishes; it remains zero in all region $p<\protect%
\gamma -1$. M$_i$ is the inversion point where $p=\protect\gamma $ and where
the acoustic flow changes sign. The full line corresponds to $\protect\gamma %
=0.4$, the broken line to $\protect\gamma =0.6$. Here
$\protect\zeta =0.8$.} \label{p-u-relation}
\end{figure}

In a quasi static regime, the flow $U$ entering the resonant
cavity is modelled with the help of an approximation of the
Bernoulli equation, as discussed e.g. in \cite{Hirschberg:90}.\ We
note $P_{int}$ the acoustic pressure inside the mouthpiece,
assumed to be equal to the one at the output of the reed channel,
$P_{m}$ the pressure inside the mouth of the player; for small
values of the difference:
\begin{equation}
\Delta P=P_{m}-P_{int}\text{ ,}  \label{delta-P}
\end{equation}
the reed remains close to its equilibrium position, and the conservation of
energy implies that $U$ is proportional to $\eta _{p}\sqrt{\left\vert \Delta
P\right\vert }$, where $\eta _{p}=\pm 1$ is the sign of $\Delta P$ (we
ignore dissipative effects at the scale of the flow across the reed
channel); for larger values of this difference, the reed moves and, when the
difference reaches the closure pressure $P_{c}$, it completely blocks the
flow.\ These two effects are included by assuming that if $\Delta P\leq
P_{c} $ the flow $U$ is proportional to $\eta _{p}\sqrt{\left\vert \Delta
P\right\vert }\left[ P_{c}-\Delta P\right] $, and if $\Delta P>P_{c}$, the
flow vanishes.\ Introducing the dimensionless quantities:
\begin{equation}
\begin{array}{l}
p=P_{int}/P_{c} \\ u=UZ_{\infty }/P_{c} \\ \gamma
=P_{m}/P_{c}\text{ } \\ \Delta p=\Delta P/P_{c}=\gamma -p\text{ .}
\end{array}
\label{defs}
\end{equation}
where $Z_{\infty }=\rho c/S$ is the acoustic impedance of an infinitely long
cylindrical resonator having the same cross section $S$ than the clarinet
bore ($\rho $ is the density of air, $c$ the velocity of sound), we obtain:
\begin{eqnarray}
u &=&0  \label{1} \\ &&\text{if \ \ \ }\Delta p>1\text{ \ \ i.e. \
\ \ \ \ \ }p<\gamma -1\text{ \ ;} \notag \\ u &=&\zeta (p+1-\gamma
)\sqrt{\gamma -p}  \label{1bis} \\ &&\text{if }0<\Delta p<1\text{
i.e. \ }\gamma -1<p<\gamma \text{ ;}  \notag \\ u &=&-\zeta
(p+1-\gamma )\sqrt{p-\gamma }  \label{1ter} \\ &&\text{if \ \ \
}\Delta p<0\text{ \ i.e. \ \ \ \ \ \ }p>\gamma .  \notag
\end{eqnarray}
The parameter $\zeta $ characterizes the intensity of the flow and
is defined as:
\begin{equation}
\zeta =\frac{cS_{op}}{S}\sqrt{\frac{2\rho }{P_{c}}}\text{ ,}  \label{zeta}
\end{equation}
where $S_{op}$ is the opening cross section of the reed channel at
rest.\ One can show
that $\zeta $ is inversely proportional to square root of the reed stiffness%
\footnote{
the reed remains close to its equilibrium position; the acoustic
flow is then independent of the stiffness of the reed.\ Equation
(\ref{1bis}) then
provides $UZ_{\infty }/P_{c}\simeq \zeta \sqrt{(P_{m}-P_{int})/P_{c}}$, or $%
U\simeq \left( \zeta \sqrt{P_{c}}/Z_{\infty }\right) \sqrt{(P_{m}-P_{int})}$%
; but $P_{c}$ is roughly proportional to the reed stiffness, so
that the
independence of the flow with respect to the stiffness requires that $\zeta $%
\ is inversely proportional to the square root of this
stiffness.}, contained in $P_{c}$. In real instruments, typical
values of the parameters are $\gamma \in \left[ 0,1.5\right] $; $\zeta =%
\left[ 0.1,0.5\right] $ ; values $\zeta >1$ will not be considered
here, since they correspond to multi-valued functions $u(p)$, a
case that does not seem very realistic in practice.
 Fig.\ref{p-u-relation}  shows an example of function defined
  in Eqs.(\ref{1} to \ref{1ter}). It is obviously non-analytic; it is
made of three separate analytic pieces, with a singular point at
$p=\gamma. $ The derivative of the function $u(p)$ is
discontinuous at $p=\gamma -1$ (point  $M_{b}$ in Fig.
\ref{p-u-relation}, the index $b$ being used for the limit of
possible beating); a smoothing of the resulting angle of the
function could easily be introduced at the price of a moderate
mathematical complication, but this is not necessary for the
present discussion.

\subsection{Iteration}\label{iteration}

Waves are assumed to be planar in the quasi one dimensional
cylindrical resonator.\ Any wave can be expanded into an outgoing
wave $p^{+}(t-z/c)$ and an incoming wave $p^{-}(t+z/c)$, where $t$
is the time and $z$ the abscissa coordinate along the axis of the
resonator; at point $z=0$ (at the
tip of the reed), the acoustic pressure and flow\footnote{%
The flow is related to the pressure via the Euler equation:
$\partial P/\partial z=-\rho S^{-1}\partial U/\partial t.$} are
given by:
\begin{equation}
p(t)=p^{+}(t)+p^{-}(t)\text{ \ \ ; \ \ }u(t)=p^{+}(t)-p^{-}(t)  \label{4b-1}
\end{equation}
or:
\begin{equation}
p^{+}(t)=\frac{1}{2}\left[ p(t)+u(t)\right] \text{ \ \ ; \ \ }p^{-}(t)=\frac{%
1}{2}\left[ p(t)-u(t)\right] .  \label{4b-2}
\end{equation}
We will use variables $p^{\pm }(t)$ instead of $p(t)$ and $u(t)$. If we
assume that the impedance at the output of the resonator is zero (no
external radiation, the output pressure remains the atmospheric pressure),
we obtain the reflection condition:
\begin{equation}
p^{-}(t)=-p^{+}(t-2\ell /c)\text{ ,}  \label{4}
\end{equation}
where $\ell $ is the resonator length and $c$ the sound velocity. This
equation expresses that the reflected wave has the same amplitude than the
incoming wave.\ Losses are not included in this relation, but one can also
introduce them very easily by replacing (\ref{4}) by:
\begin{equation}
p^{-}(t)=-\lambda p^{+}(t-2\ell /c)\text{ ,}  \label{4c}
\end{equation}
which amounts to introducing frequency independent losses; a typical value
is $\lambda =0.9$. For a cylindrical, open, tube with no radiation at the
open end so that losses only occur inside the tube, $\lambda =\exp (-2\alpha
\ell )$, where $\alpha $ is the absorption coefficient.\ Of course this is
an approximation: real losses are frequency dependent\footnote{%
The value of $\alpha $ depends on both frequency $f$ and radius $R.$ For
normal ambient conditions ($20%
{{}^\circ}%
C$), $\alpha =2.96\;10^{-5}\sqrt{f}/R$ (see e.g. \cite{causse}).} and
radiation occurs but, since losses remain a relatively small correction in
musical instruments, using Eq. (\ref{4c}) is sufficient for our purposes.

We now assume that all acoustical variables vanish until time $t=0$, and
then that the excitation pressure in the mouth suddenly takes a new constant
value $\gamma $; this corresponds to a Heaviside step function for the
control parameter.\ Between time $0$ and time $2l/c$, according to (\ref{4c}%
), the incoming amplitude $p^{-}(t)$ remains zero, but the outgoing
amplitude $p^{+}(t)$ has to jump to value $p_{1}^{+}$ in order to fulfil Eqs.(\ref{1} to \ref{1ter})
.\ At time $t=2l/c$, the variable $p^{-}(t)$ jumps to value $%
-\lambda p_{1}^{+}$, which immediately makes $p^{+}(t)$ jump to a new value $%
p_{2}^{+}$, in order to still fulfil Eqs.(\ref{1} to \ref{1ter}).\
This remains true until
time $t=4l/c$, when $p^{-}(t)$ jumps to value $-\lambda p_{2}^{+}$ and $%
p^{+}(t)$ to a value $p_{3}^{+}$, etc. By recurrence, one obtains a regime
where all physical quantities remain constant in time intervals $%
2nl/c<t<2(n+1)l/c$, in particular $p_{n}$ for the pressure and
$u_{n}$ for the flow, with the recurrence relation:
\begin{equation}
p_{n}^{-}=-\lambda p_{n-1}^{+}\text{.}  \label{recurrence}
\end{equation}
In what follows, it will be convenient to use $2l/c$ as a natural time
unit.\ We will then simply call \textquotedblleft time $n$%
\textquotedblright\ the time interval $(n-1)2l/c\leq t<n2l/c$.
Notice that in order to get higher regimes (with e.g. triple
frequency), the previous choice of transient for $\gamma$ needs to
be modified (see e.g.\cite{Kergomard:95}).

Now, by combining Eqs.(\ref{1} to \ref{1ter}) and \ref{4b-1}), one
can obtain a non-linear relation $g$ between $p_{n}^{+}$ and
$p_{n}^{-}$:
\begin{equation}
p_{n}^{+}=g(p_{n}^{-})\text{ ,}  \label{eqnp}
\end{equation}
which, combined with (\ref{recurrence}), provides the relation:
\begin{equation}
p_{n}^{+}=g(-\lambda p_{n-1}^{+})=f(p_{n-1}^{+})\text{,}
\label{def-fonction}
\end{equation}
with, by definition: \ \ $f(x)\equiv g(-\lambda x).$
 The equation of evolution of the
system are then equivalent to a simple mapping problem with an
iteration function $f(x)$. The graph of this function is obtained
by rotating the non-linear characteristics of Fig. \ref
{p-u-relation} by $45^{\circ }$ (in order to obtain $g$), then
applying a symmetry (to include the change of sign of the
variable) and finally a horizontal rescaling by a factor
$1/\lambda $; the result is shown in Fig.\ \ref{fonction-iteree}.\
This provides a direct and convenient graphical construction of
the evolution of the system \cite{Maganza:86}; Fig. \ref
{iteration} shows how a characteristic point $1$ is transformed
into its next iterate $2,$ etc... by the usual construction, at
the intersection of a straight line with the iteration curve, i.e.
by transferring the value of $f(x)$ to the $x$ axis and reading
the value of the function at this abscissa in order to obtain
$f\left[ f(x)\right] $.

In what follows, we consider $\gamma $ as the main control parameter of the
iteration; it corresponds to a change of pressure in the mouth of the
instrumentalist. A\ second control parameter is $\zeta $, which the player
can also change in real time by controlling the lip pressure on the reed.\
For a given note of the instrument, parameter $\lambda $ remains fixed, but
of course depends on which lateral holes of the clarinet are closed, in
other words on the pitch of the note.\

The oscillations where the functions remain constant and jump to a different
value at regular interval of times are reminiscent of the Raman regime for
the oscillation of bowed strings \cite{raman}. Mc Intyre et al. have indeed
noticed that, if one replaces the non-linear function by that corresponding
to a bowed string, one obtains the Raman oscillation regime of a string
bowed at its center \cite{McIntyre:83}.

\section{Properties of the iteration}

\label{properties}

The analytical expression of the iteration function is given in Appendix \ref
{app1}. Figure \ref{fonction-iteree} shows the function for given values of
the parameters $\gamma $ and $\zeta $, and three different values of the
loss parameter $\lambda .$

\begin{figure}[h]
\centering\includegraphics[width=10cm]{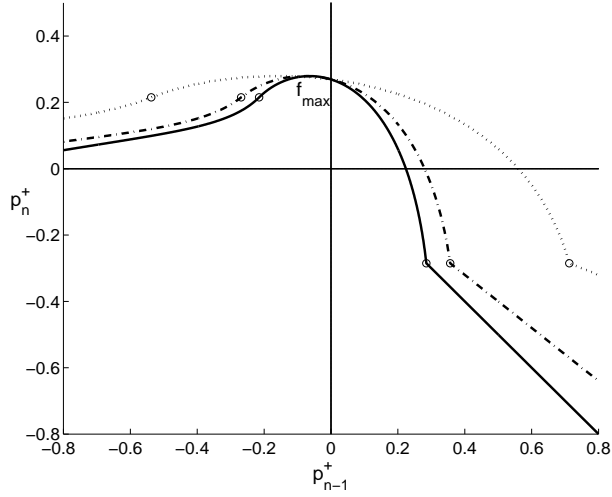}
\caption{Iteration function $f$ for $\protect\zeta =0.8$ and $\protect\gamma =0.43$ \ Solid line $%
\protect\lambda =1$ (no loss in the resonator); mixed line $\protect\lambda %
=0.8$; dotted line $\protect\lambda =0.4$. The circles on the right indicate
the contact point M$_{b}$, those on the left the flow inversion point M$_{i}$.%
}
\label{fonction-iteree}
\end{figure}

In the literature, the most commonly studied functions have the following
properties (see e.g. Collet and Eckman \cite{Collet-Eckmann} or Berg\'{e} et
al \cite{Berge-Pomeau-Vidal}):

\begin{itemize}
\item  They are defined on a finite interval and map this interval into
itself;

\item  They are continuous;

\item  They have a unique maximum;

\item  their Schwarzian derivative is negative.
\end{itemize}

A function verifying these properties will be called a \textquotedblleft
standard\textquotedblright\ function; the function $f(x)$ of interest in our
case does not fulfil all these requirements.

\subparagraph{Domain of iteration}

\label{domain}

Usually, the iteration function defines an application of the interval $%
0\leq x\leq 1$ over itself.\ Here, $f(x)$ is defined on an
infinite interval $\left[ -\infty ,+\infty \right] $ even if,
obviously, very large values of the variables are not physically
plausible. Nevertheless, analyzing the different cases
corresponding to Eqs.(\ref{1} to \ref{1ter}), one can show that
the function\ $f(x)$ has a maximum $f_{\max }$ obtained for:
\begin{equation}
x_{\max }=-\frac{1}{\lambda }\left[ \frac{\gamma }{2}-\frac{5}{18}-\chi
\left( \zeta -\frac{5}{3\zeta }\right) \right]  \label{xmax}
\end{equation}
with value:
\begin{equation}
f_{\max }=\frac{\gamma }{2}+A_{\zeta }\text{ with \ }A_{\zeta }=\chi \left(
\zeta +\frac{1}{3\zeta }\right) -\frac{1}{18}\text{ }  \label{5A}
\end{equation}
where $\chi $ is defined by:
\begin{equation}
\chi =\frac{1}{9}\left[ \sqrt{3+\frac{1}{\zeta ^{2}}}-\frac{1}{\zeta }\right]
\text{ .}  \label{5B}
\end{equation}

It can be shown that this maximum is unique for large value of $\zeta $ ($%
\zeta >1/\sqrt{3})$; for smaller values, a second maximum exists at a very
large negative values of $x$, i.e. for very large negative flow, but we will
see below that such values of the flow cannot be obtained after a few
iterations.\ Therefore we focus our attention only on the maximum $f_{\max }$%
, which varies slowly as a function of $\zeta $ because $A_{\zeta }$
increases monotonically from 0 for $\zeta =0$ to a small value ($5/54,$ for $%
\zeta =1$).

The geometrical construction of Fig. \ref{iteration} shows that,
after a single iteration, the characteristic point M necessarily
falls at an abscissa $x\leq f_{\max }$.\ Let us call $f_{\min
}=f(f_{max})$ the ordinate of the point on the iteration function
with abscissa $f_{\max }$.\ The two vertical lines $x=f_{\min }$
and $x=f_{\max }$, together with the two horizontal lines
$y=f_{\min }$ and $y=f_{\max }$, define a square in the $x,y$
plane, from which an iteration cannot escape as soon as the
iteration point
has fallen inside it \footnote{%
We assume that $f(f_{min})>f_{min}$, which means that the iteration curve
crosses the left side of the square, as is the case in Fig \ref{iteration}.}%
.\ Conversely, since every characteristic point has at least two
antecedents, the iteration can bring a point that was outside the
square to inside.\ In other words, the square determines a part of
the curve which is invariant by action of the function.\ For usual
initial conditions, such as $p_{0}^{-}=0$, the starting point
already lies within the square, so that all points of the
iteration keep this property. We have checked that, even if one
starts with very large and unphysical pressure differences
(positive or negative), the iterations rapidly converge to the
inside the square. In what follows, we call it the
\textquotedblleft iteration square\textquotedblright .

The net result is that, if we do not consider transients, we can consider
that the function defines an application of the interval $\left[ f_{\min },%
\text{ }f_{\max }\right] $ over itself.\ We are then very close to the usual
mapping situation, except that here the interval depends on the control
parameters (since the value of $f_{\max }$ depends on $\gamma $ and $\zeta $%
), but with a relatively slow variation.
\begin{figure}[h]
\centering \includegraphics[width=10cm]{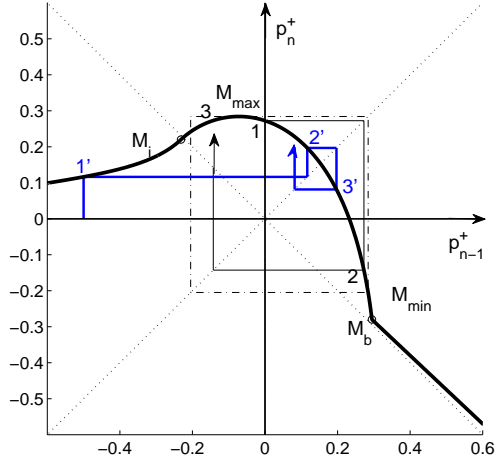}
\caption{Graphical illustration of the iteration, where an initial
point 1 is iterated into point 2, 3, etc. (similarly for the point
1', 2', 3',...). Since the non-linear iteration function has a
maximum $f_{max}$, after a few steps the iteration remains inside
an ``iteration square" shown in broken lines. This square has its
upper side tangent to the maximum of the function, at point
$M_{max}$, which after one iteration becomes point $M_{min}$
defining the lowest side of the square (ordinate $f_{min}$).
Depending on the parameters, the iteration square contains or does
not contain the contact point $M_{b}$ and the flow inversion point
$M_{i}$. Here $\protect\gamma =0.44$, $\protect\zeta=0.8$,
$\protect\lambda =0.95$.} \label{iteration}
\end{figure}

\subparagraph{Singularities}

An interesting feature of the iteration function is the
discontinuity of its first derivative occurring at the beating
limit point $M_{b}$ at $x=x_{b}$, given by:
\begin{equation}
x_{b}=\frac{1-\gamma }{2\lambda }\text{ \ \ \ ; \ \ \ }f(x_{b})=\frac{\gamma
-1}{2}.  \label{5b}
\end{equation}
When the reed closes the channel ($p^{+}=p^{-}$, $p^{+}+p^{-}<\gamma -1$), $%
x>x_{b}$, $\ f(x)=-\lambda x$, the iteration function is linear.

Another singularity, i.e. a discontinuity of the second derivative, is
obtained at the crossover between positive and negative flow, the inversion
point $M_{i}$ where sign of the flow changes.\ Its abscissa $x=x_{i}$ is
given by:
\begin{equation}
x_{i}=-\frac{\gamma }{2\lambda }\text{ \ \ \ ; \ \ \ }f(x_{i})=\frac{\gamma
}{2}.  \label{5c}
\end{equation}
For $0<\gamma <1$, $x_{i}$ is negative and $x_{b}$ positive: therefore the
initial point of the iteration ($x=0$) lies in the interval $\left[
x_{i},x_{b}\right] $, with neither contact with the mouthpiece nor negative
flow, as one could expect physically.

\subparagraph{Schwarzian derivative}

The Schwarzian derivative \cite{Collet-Eckmann} of $f(x)$ is equal to:
\begin{equation}
Sf=\frac{f^{\prime \prime \prime }}{f^{\prime }}-\frac{3}{2}\left[ \frac{%
f^{^{\prime \prime }}}{f^{\prime }}\right] ^{2}\text{,}
\end{equation}
where $f^{\prime }$, $f^{\prime \prime }$ and $f^{\prime \prime
\prime }$ indicate the first, second and third derivatives of
$f(x)$, respectively. If $x>x_{b}$, it is zero; if
$x_{i}<x<x_{b}$, using the change of variables given in Appendix
\ref{app1}, $Sf$ can be shown to be equal to:
\begin{equation}
Sf=\frac{8\lambda ^{2}}{Y^{^{\prime }4}(Y^{\prime }-2)^{2}}\left[ Y^{\prime
\prime \prime }Y^{\prime }(Y^{\prime }-2)-3Y^{\prime \prime 2}(Y^{\prime }-1)%
\right] ,
\end{equation}
where $Y$ is a function of $X$ - see Eqs. (\ref{a1}) to (\ref{a3}).
Therefore its sign does not depend on the loss parameter $\lambda $. After
some calculations, the Schwarzian derivative is found to be negative for all
$x\in \left[ x_{i},x_{b}\right] $ when $\zeta <1/\sqrt{5}.$ Otherwise, it is
negative up to a certain value, then positive up to $x=x_{b}.$ The
calculation of $Sf$ for the case $x<x_{i}$ shows that it is positive, except
for a small interval. The iteration function therefore differs from a
standard function because of the sign of the Schwartzian derivative; this is
related to the nature of the bifurcation at the threshold of oscillation
\cite{mayer}, which can be either direct or inverse.

\subparagraph{Beating and negative flow limits \label{beating-limits}}

In Fig. \ref{iteration} we see that, depending whether the contact point $%
M_{b}$ \ and flow inversion point $M_{i}$ of the iteration curve fall inside
or outside the iteration square, a beating behavior of the reed and a sign
inversion of the air flow are possible or not.

Point $M_{b}$ falls inside the iteration square if its abscissa $x_{b}$
given by (\ref{5b}) is smaller than $f_{max}$, which leads to:
\begin{equation}
\gamma >\gamma _{b}\equiv \frac{1-2\lambda A_{\zeta }}{1+\lambda }.
\label{5d}
\end{equation}
\begin{figure}[h]
\centering \includegraphics[width=10cm]{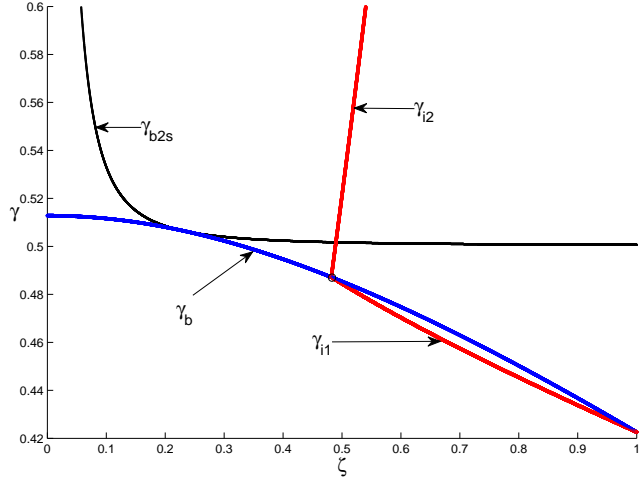}
\caption{In the plane of the control parameters $\protect\gamma $ and $%
\protect\zeta $, the line representing $\protect\gamma _{b}$ gives
a border between the upper region, where reed beating may occur,
and the lower region where it cannot - see Eq. (\ref{5d}).\ As a
point of comparison, the line labelled $\protect\gamma _{b2s}$
correspond to the limit obtained in \protect\cite{Dalmont:05} for
the particular case of a 2-state regime, and given by
Eq.(\ref{5f}). The figure also shows the line $\protect\gamma
_{i1}$ and $\protect\gamma _{i2}$ associated with the possibility
of negative flow (the first one turns out to be very close to that
associated with reed
beating). Small values for the losses have been assumed ($\protect%
\lambda =0.95$).}
\label{limits}
\end{figure}
The limiting value $\gamma _{b}$ is less than unity (it tends to $1/2$ when $%
\zeta $ tends to 0 and $\lambda $ tends to unity). This necessary
condition is completely independent of the nature of the limit
cycle, and less stringent than the limit\ $\gamma _{b2s}$ obtained
in \cite{Dalmont:05}, for a 2-state cycle :
\begin{equation}
\gamma _{b2s}=1/2\left[ 1+\beta \beta _{1}+\beta _{1}^{2}(1-\beta \beta _{1})%
\right] ,  \label{5f}
\end{equation}
where $\beta =\zeta (1-\lambda )/(1+\lambda )$ and $\beta
_{1}=\beta /\zeta ^{2}.$ Fig. \ref{limits} gives a comparison
between the two limits.
  Similarly, a
necessary condition for possible inversions of the sign of the air
flow is that point $M_{i}$ falls inside the iteration square of
Fig \ref {iteration}, in other words that $x_{i}$ is larger than
$f_{\min }$.\ We show in Appendix \ref{app2} that this happens if:
\begin{equation}
\gamma _{i1}<\gamma <\gamma _{i2}.  \label{10j}
\end{equation}
The expression of the two limits $\gamma _{i1}$ and $\gamma _{i2}$ are given
in the Appendix and can be seen on Fig. \ref{limits}. \ They are solutions
of $x_{i}=f_{\min }$, and exist only if the following condition holds:
\begin{equation}
\lambda >\frac{1}{1+2A_{\zeta }}.  \label{10h}
\end{equation}
\ Therefore, for a given $\zeta ,$ negative flow is possible only above a
certain value of $\lambda $; this value is $27/32=0.84$ for $\zeta =1$, and
tends to unity when $\zeta $ tends to zero. Using a more realistic shape for
the function $f(x)$ with a rounding of the kink at $x_{b}$ (no discontinuity
of the derivative) should lead to a shorter range of negative flow, making
the phenomenon even less likely, as illustrated in Fig.\ \ref{limits}.

Of course, the two above conditions (\ref{5d}) and (\ref{10j}) are
necessary, but not sufficient; they do not ensure that either beating or
flow inversion will indeed take place, since this will be true only if the
corresponding regions of the non-linear curve are reached during the
iterations.\ Generally speaking, this will have more chance to occur in
chaotic regimes, where many points are explored in the iterations, than in
periodic regimes. Since no observation of negative flow has been reported in
the literature, it is not clear whether this actually happens in real
instruments.

In conclusion of this section, the iteration function is similar to those
usually considered in the context of iterated maps, without really belonging
to the category of \textquotedblleft standard\textquotedblright\ functions.\
The major difference is actually the effect of the control parameters on the
function, since usually the control parameters acts as a gain, expanding the
vertical axis of the graph; here the parameter $\gamma $ (pressure in the
mouth of the instrumentalist) translates the iteration function along an
axis at $45^{\circ }$ of the coordinate axis, while the other control
parameter $\zeta $ (the pressure of the lip on the reed) expands the
function along the perpendicular axis.\ It is therefore not surprising that
we should find a parameter dependence of the dynamical behaviors that is
significantly different from the standard results.\

\section{ Bifurcation curves}

\label{bifurcations}

\begin{figure}[h]
\centering \includegraphics[width=10cm]{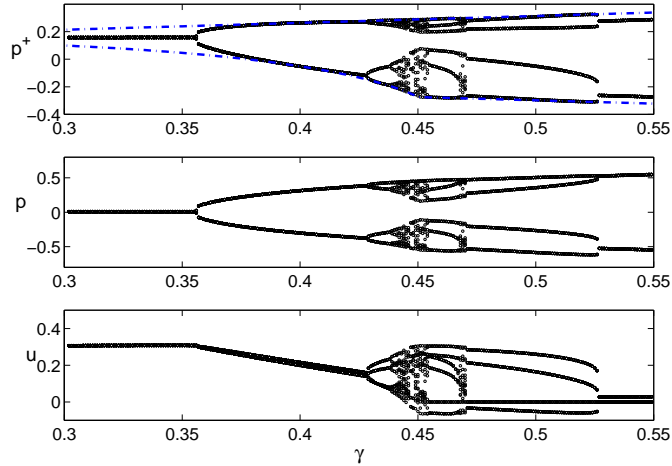}
\caption{{}Bifurcation curve for $\protect\lambda =0.95$ and $\protect\zeta %
=0.8$. For increasing values of the abscissa $\protect\gamma $
(blowing pressure), corresponding to a crescendo, the curve show
the values after 400 iterations of the outcoming wave $p^{+}$
(top) the pressure $p$ in the mouthpiece (middle), and the volume
flow $u$ (bottom). Above $\protect\gamma \simeq 0.45$, the flow
can be negative and the reed
can beat.\ The top figure also shows $f_{\max }(\protect\gamma )$ and $%
f_{\min }(\protect\gamma )$ (mixed lines) associated to the
\textquotedblleft iteration square\textquotedblright .}
\label{fig-bif}
\end{figure}

Figure \ref{fig-bif} shows an example of bifurcation curves, for $\lambda
=0.95$ and $\zeta =0.8$, and illustrates the relative complexity of the
possible regimes. The upper curve corresponds to the outgoing amplitude $%
p^{+}$ (or $x$), the middle curve to acoustic pressure $p$, and
the lower curve to the acoustic flow $u$. The three curves show
the last 20 values obtained after computing 400 iterations for
each value of the mouth pressure $\gamma $. By calculating $2000$
iterations for a given value of the parameter $\gamma $, we have
checked that the limit cycle is then reached. Obviously this
method leads to stable regimes only.

When the control parameter $\gamma $ increases, the beginning of
these curves follows a classical scenario of successive period
doublings, leading eventually to chaos; as expected, high values
of the parameters $\zeta $ and $\lambda $ favour the existence of
chaotic regimes, as well as beating reed or negative flow.\ When
$\gamma $ continues to increase, another phenomenon takes place:
chaos disappears and is replaced by a reverse scenario containing
a series of frequency (instead of period) doublings.\ We call this
phenomenon a \textquotedblleft backwards
cascade\textquotedblright\ (in order to distinguish it from the
usual \textquotedblleft inverse cascade\textquotedblright , which
takes place within periodicity windows inside chaos
\cite{Berge-Pomeau-Vidal}); this backwards cascade is a
consequence of the specificities of the effect of the control
parameter on the iteration function in our model, and of the
particular shape of the iteration function (for instance a
straight line beyond the beating limit point). As a matter of
fact, different kinds of cascades have been studied in the
literature (see e.g. \cite{lau1} and \cite{lau2}, in particular
Fig.5).\\
  In Fig. \ref{fig-bif}, the
variations of $\gamma $ correspond to a \textquotedblleft
crescendo\textquotedblright : for a given value of $\gamma $, the
initial value for the iteration, $p_{0}^{+}$, is chosen to be
equal to the last value $p_{400}^{+}$ obtained with the previous
value of $\gamma $. But we have also studied the \textquotedblleft
decrescendo\textquotedblright\ regime and observed that, in the
chaotic regimes, the plotted points differ from the crescendo
points; on the other hand, they remain the same in the periodic
regimes, indicating a direct character of the bifurcations (no
hysteresis). We have found an exception to this rule: between
values $\gamma =0.5$ and $\gamma =0.53$, 2-state and 4-state
regimes coexist, indicating an inverse bifurcation. Another
inverse bifurcation, between a 2-state regime and a static regime,
occurs beyond the limit of the figure, the two regimes coexisting
between $\gamma =1$ and $\gamma =6.3544$; this is not shown here
(the shape of the curve can be found in Ref. \cite{Dalmont:05},
see upper Fig. 4).

The two limits of the function $f(x)$, $f_{\max }$ and $f_{\min
},$ are also plotted in the upper figure ( $p^{+}$) showing that,
as expected, the corresponding values remain inside the iteration
square (\S\ \ref{domain}). In the figure at bottom, the results
for the flow $u$ exhibit lower limits for negative flow and for
beating, which are very close to the theoretical limits,
respectively $\gamma _{i1}=0.4454$ and $\gamma _{b}=0.4503,$ and
are located within the chaotic regime.\ Negative flow disappears
at the bifurcation between the 4-state and the 2-state regime,
$\gamma =0.5262$, a
much lower value than the higher limit for negative flow $\gamma _{i2}=1.189$%
.

\begin{table}[tbp]
$
\begin{tabular}{|l|l|l||l|l|l|}
\hline From $\gamma =$ & Regime & Comments & From $\gamma =$ &
Regime & Comments \\ \hline $0$ & 1-state &  & $0.4540$ & 24-state
&  \\ \hline $0.3545$ & 2-state &  & $0.4542$ & 12-state &
\\ \hline $0.4272$ & 4-state &  & $0.4544$ & 6-state &  \\
\hline $0.4384$ & 8-state &  & $0.466216$ & I &  \\ \hline
$0.4403$ & 16-state &  & $0.4664$ & chaos &  \\ \hline $0.4408$ &
32-state & & $0.46945$ & 60-state &  \\ \hline $0.4409$ & chaos &
$\gamma _{i1}=0.4454$ & $0.4695$ & 20-state &  \\ \hline $0.4467$
& 6-state & PW & $0.4696$ & chaos &  \\ \hline $0.4474$ & 12-state
& PW & $0.46985$ & 4-state &  \\ \hline $0.4476$ & 24-state & PW &
$0.5000$ & 2-state (D) &  \\ \hline $0.4479$ & chaos & $\gamma
_{b}=0.4503$ & $0.53$ & 2-state (C) &  \\ \hline $0.4538$ &
36-state &  & $1.$ & 1-state (D) & $\gamma _{i2}=1.189$
\\ \hline $0.4539$ & chaos &  & $6.3544$ & 1-state (C) &  \\
\hline
\end{tabular}
$%
\caption{Values of the parameter $\protect\gamma $ at the
\textit{lower} limit of the different regimes, corresponding to
Fig. \ref{fig-bif}. I=intermittencies; C= crescendo; D=
decrescendo; PW= periodicity windows.}
\end{table}


Table 1 shows the critical values of $\gamma $ corresponding to changes of
regime. Up to the first chaotic regime ($\gamma =0.4409$), the behavior
follows the usual period doubling cascade scenario. Between $\gamma =0.4467$
and $\gamma =0.4479$ a \textquotedblleft periodicity
windows\textquotedblright\ \cite{Berge-Pomeau-Vidal} is obtained, with
6-state, then 12-state and 24-state regimes (but no 3-state regime). Above
the value $\gamma =0.4409$ for which chaos starts, an \textquotedblleft
inverse cascade\textquotedblright\ type scenario is observed, then
intermittences occur, chaos again, and finally the \textquotedblleft
backwards cascade\textquotedblright\ to the static regime. We did not try to
obtain the same accuracy for the values of all different thresholds, because
the ranges for $\gamma $ have very different widths; for some values of $%
\gamma $, it has been necessary to make up to 2000 iterations, and sometimes
it is not obvious to distinguish between a chaotic regime, a long transient,
or an intermittency regime.

\section{ Iterated functions}

\label{iterated}

We now discuss how the iterated functions can be used to study the different
regimes and their stability. We write $f^{(2)}(x)$ the second iterate of $f$%
, and more generally $f^{(n)}(x)$ its iterate of order $n$; the derivative
of $f$ with respect to $x$ is $f^{\prime }(x)$. Around the fixed point $%
x^{\ast } $ of the first iterate $f(x)$, a Taylor expansion gives:
\begin{equation*}
f(x)=f(x^{\ast })+(x-x^{\ast })f^{\prime }(x^{\ast })+..=x^{\ast
}+(x-x^{\ast })f^{\prime }(x^{\ast })+..\text{ \ \ ,}
\end{equation*}
which provides the well known stability condition for a fixed point $x^{\ast
}$ of $f(x)$:
\begin{equation}
\left\vert f^{\prime }(x^{\ast })\right\vert <1. \label{eee}
\end{equation}
Since the derivative of the iterate of order $n$ is given by:
\begin{equation*}
f^{(n)\prime }(x^{\ast })=f^{\prime }(x^{\ast })f^{(n-1)\prime }(x^{\ast })
\end{equation*}
one can show by recurrence that, when $x=x^{\ast}$, it is equal to
the $n$-th power of the derivative of $f$, so that:
\begin{equation}
f^{(n)}(x)=x^{\ast }+(x-x^{\ast })\left[ f^{\prime }(x^{\ast
})\right] ^{n}+.. \label{ddd}
\end{equation}
If the fixed point is stable (resp. unstable) with respect to $f(x)$, it is
also stable (resp. unstable) with respect to any iterate. If $x$ is a
vector, instead of a scalar, this linearized approach leads to the Floquet
matrix, and $f^{\prime }(x)$ should be replaced by the eigenvalues of the
matrix.

\subsection{Stability of the period doubling regimes}

\begin{figure}[h]
\centering \includegraphics[width=10cm]{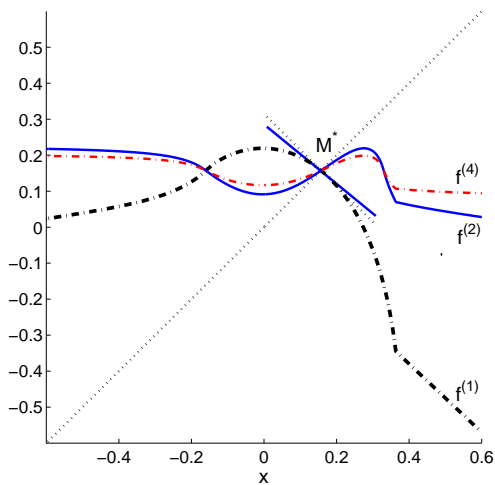}
\caption{{}Iteration functions for $\protect\lambda =0.95,$ $\
\protect\gamma =0.31$ and $\protect\zeta =0.8$. The 1st iterate
$f(x)$ is shown with a mixed line, the 2nd iterate $f^{(2)}(x)$
with a (blue) solid line, and $f^{(4)}(x)$ with a (red), thin
mixed line. The dotted lines are the first diagonal and the
straight line perpendicular at the fixed point $M^{\ast }$ of
$f(x)$,
solution of $f(x)=x.$ The tangent lines to iterate 1 at the point $%
M^{\ast }$ is shown with a solid line.} \label{iteree_gamma=031}
\end{figure}
\begin{figure}[h]
\centering \includegraphics[width=10cm]{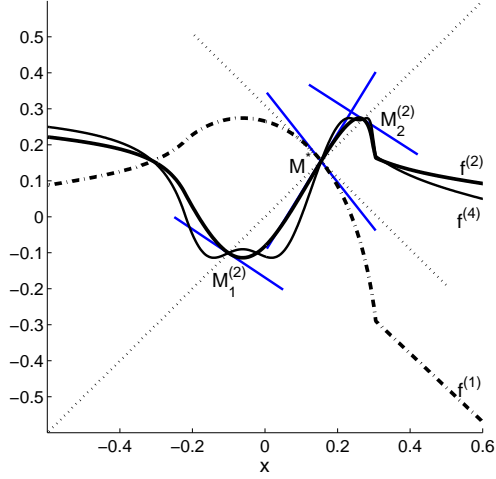}
\caption{Iterates for $\protect\lambda =0.95,$ $\ \protect\gamma
=0.42$ and $\protect\zeta =0.8$, with the same plots that in
figure \ref
{iteree_gamma=031}. The tangent lines at the new fixed points of $f^{(2)}(x)$%
, $M_{1}^{(2)}$ and $M_{2}^{(2)}$, are also shown.}
\label{iteree_gamma=042}
\end{figure}
 Examples of iterated functions
of order 1, 2, and 4 are shown in figures \ref
{iteree_gamma=031} and \ref{iteree_gamma=042}, with the same values of $%
\zeta $ and $\lambda $ as in figure \ref{fig-bif}; in the former, the
blowing pressure $\gamma $ is $0.31$, in the latter, $\gamma $ is $0.42$.\ \
The first iterate has a unique fixed point, $M^{\ast }=(x^{\ast }$, $x^{\ast
})$, located by definition on the first diagonal.\ The fixed point is stable
if the absolute value of the derivative at $M^{\ast }$ is smaller than
unity, in other words if the tangent line lies between the first diagonal
(with slope $+1$) and its perpendicular (with slope $-1$).\ When $\gamma =$ $%
0.31$, we see in Fig.\ \ref{iteree_gamma=031} that the fixed point
$M^{\ast } $ is stable, so that no oscillation takes place. When
$\gamma$ increases, M* becomes instable and, at the same time,
gives rise to three fixed points of $f^{(2)}$. \ For $\gamma
=0.42$, Fig.\ \ref{iteree_gamma=042} shows that the tangent is
outside the angle between the diagonal and its perpendicular, so
that the fixed point is now unstable; on the other hand, the
second iterate $f^{(2)}$ now has two more fixed points
$M_{1}^{(2)}$ and $M_{2}^{(2)}$ with slopes less than $1$ (in
absolute value): we therefore have a stable 2-state regime.

The same scenario then repeats itself when $\gamma $ continues to increase:
at some value, points $M_{1}^{(2)}$ and $M_{2}^{(2)}$ become instable in
turn (the corresponding slope exceeds $1$ in absolute value), and both
points $M_{1}^{(2)}$ and $M_{2}^{(2)}$ divide themselves into three fixed
points of $f^{(4)}$; the two extreme new points have small slopes for this
iterate, which leads to a 4-state stable regime.\ By the same process of
successive division of fixed points of higher and higher iterates, one
obtains an infinite number of period doublings, until eventually chaos is
reached.\ This is the classical Feigenbaum route to chaos.

Some general remarks are useful to understand the shape of the iterates in
the figures:

\begin{itemize}
\item  If the value of $f(x)$ for the abscissa $x$ verifies $f(x)=f(x^{\ast
}) $, i.e. if the point $M$($x$, $f(x)$) is on a horizontal line
$y=x^{\ast } $, all iterates go through the same point;

\item  The extrema of $f^{(2)}(x)$ verify either $f^{\prime }(x)=0$ (i.e. $%
x=x_{\max }$) or $f^{(2)}(x)=f_{\max }$ , because $df^{(2)}(x)/dx=f^{\prime }%
\left[ f(x)\right] f^{\prime }(x)$; therefore the extrema of
$f^{(2)}(x)$ are at either the same abscissa or the same ordinate
as those of $f(x)$;

\item  More generally, for $n>1$, if $f^{(n-1)}(x)=x_{\max }$, then  $%
f^{(n)}(x)=f_{\max }$, and it is at a maximum (its first
derivative vanishes
and the second one is negative), and if $f^{(n-1)}(x)=f_{\max }$, then $%
f^{(n)}(x)=f_{\min }$, \ and it is at a minimum (its first
derivative vanishes and the second one is positive);

\item  The kink of the first iterate (beating limit point) is also visible on the
iterates;

\item  A well known property of the Schwarzian derivative is as follows : If the Schwarzian derivative
 of $f(x)$ is negative, the Schwarzian derivatives of all iterates are negative as well.
\end{itemize}

\begin{figure}[h]
\centering \includegraphics[width=10cm]{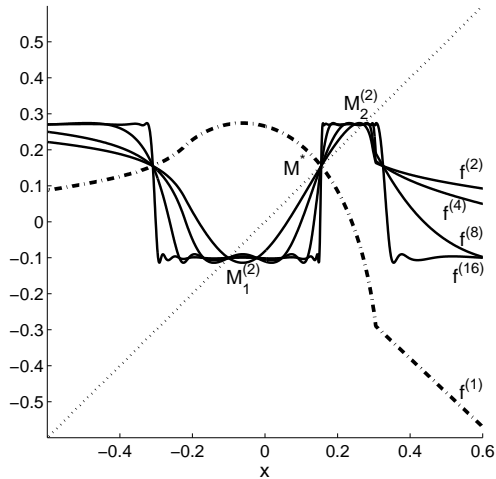}
\caption{{}Iterates for $\protect\lambda =0.95,$ $\ \protect\gamma
=0.42$ and $\protect\zeta =0.8$, of order 1, 2, 4, 8 and 16. The
convergence to the 2-state regime is visible.} \label{iteree_16}
\end{figure}
 Figure \ref{iteree_16} shows
the higher order iterates (of order 4, 8 and 16) in the same
conditions as figure \ref{iteree_gamma=042}. We observe that the
iterates become increasingly close together when their order
increases, with smaller and smaller slopes at the fixed points
corresponding to the 2-state regime. Moreover, they resemble more
and more a square function, constant in various domains of the
variable. This was expected: in the limit of very large orders,
whatever the variable is (i.e. whatever the initial conditions of
the iteration are) one reaches a regime where only two values of
the outgoing wave amplitude are possible; these values then remain
stable, meaning that the action of more iterations will not change
them anymore. So, one can read directly that the limit cycle is a
2-state on the shape of $f^{16}$, which has two values; it would
for instance have 4 in the limit cycle was a 4-state regime for
these values of the parameters. For the clarity of the figure, we
have shown only iterates with orders that are powers of $2$, but
it is of course easy to plot all iterates. For a 2-state regime,
even orders are sufficient to understand the essence of the
phenomenon, since odd order iterates merely exchange the two fixed points $%
M_{1}^{(2)}$ and $M_{2}^{(2)}$.

\begin{figure}[h]
\centering \includegraphics[width=10cm]{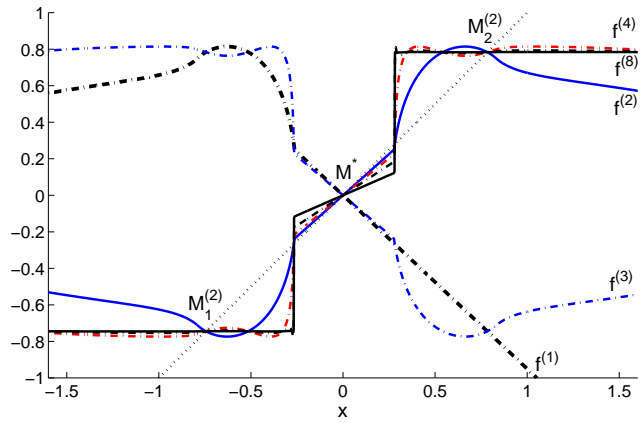}
\caption{{}Iterates for $\protect\lambda =0.95,$ $\ \protect\gamma
=1.5$ and $\protect\zeta =0.8$, of order 1, 2, 3, 4, and 8. The
 curves of $f^{(8)}$ and $f^{(16)}$ are almost perfectly
superimposed. Around $x=0$, the convergence to the static regime
appears to be very slow. On the contrary the convergence to the
2-state regime is rapid. } \label{iteree_16B}
\end{figure}

In table 1, the existence of two different stable regimes for the
same value of the parameters signals an inverse bifurcation;
Figure \ref{iteree_16B} shows an example of such a situation.\ For
$\gamma =1.2$, both the static and 2-state regimes are then
stable, depending on the initial conditions.
For the static regime, the curve $f^{(1)}$ coincides with the second diagonal $%
y=-x$, a case in which the fixed point is presumably stable (the
stability becomes intuitive when one notices that the tangents of
the higher order curves lie within the angle of the two
diagonals). For the 2-state regime, the state of positive pressure
value corresponds to a beating reed.

\begin{figure}[h]
\centering \includegraphics[width=10cm]{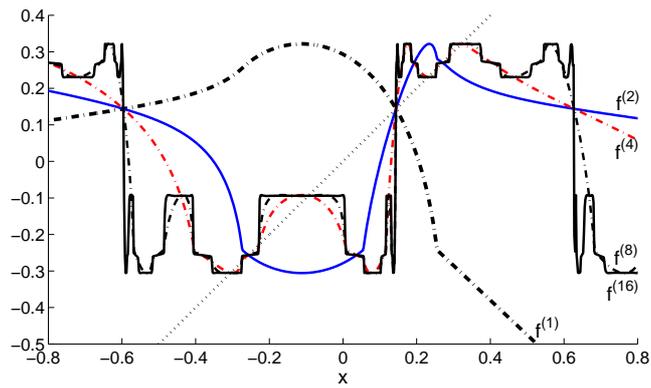}
\caption{{}Iterates for $\protect\lambda =0.95,$ $\ \protect\gamma
=0.515$ and $\protect\zeta =0.8$, of order 1, 2, 4, 8 and 16. }
\label{iteree_16C}
\end{figure}

 \clearpage
Finally Fig.\ref{iteree_16C} shows another case of existence of
two different regimes for the same value of the parameters. A
2-state regime can occur, as well as a 4-state regimes can occur.
It appears that the second one is more probable than the first
one, when initial conditions are varied.
\subsection{Periodicity windows; intermittencies}

\label{interm} We now investigate some regimes occurring in a narrow range
of excitation parameter $\gamma $.

\begin{figure}[h]
\centering \includegraphics[width=10cm]{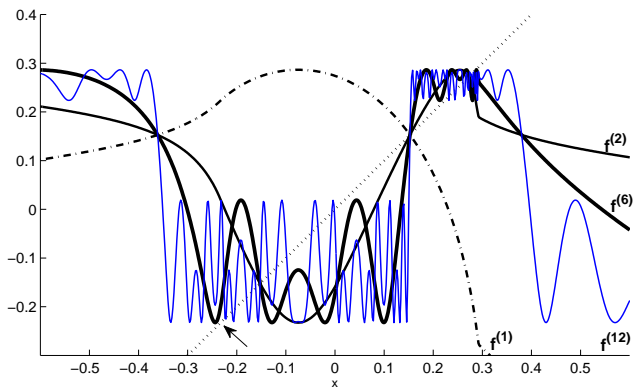}
\caption{{}Iterates for $\protect\lambda =0.95,$ $\ \protect\gamma
=0.4445$ and $\protect\zeta =0.8$, of order 1, 2, 6 and 12. A
convergence to an aperiodic regime is visible. The arrow indicates
a region where $f^{(6)}(x)$ is very close to the first diagonal,
but does not yet cross it.} \label{iteree_12A}
\end{figure}

\begin{figure}[h]
\centering \includegraphics[width=10cm]{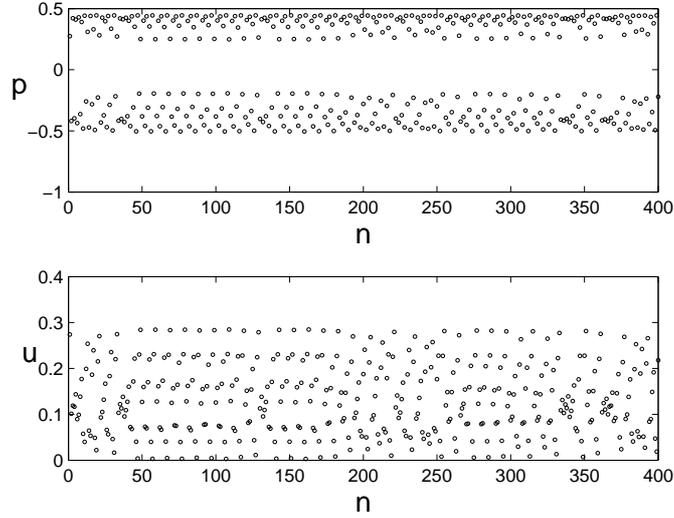}
\caption{{}Iteration from $n=0$ for $\protect\lambda =0.95,$ $\protect\zeta =0.8$, $%
\protect\gamma =0.4445$, $p_{0}^{+}=0$; the upper part shows the
the pressure $p$, the lowest part the values of the flow $u$. The
regime looks chaotic.} \label{figtransit_04445}
\end{figure}

(i) We first examine a chaotic regime occurring just before a
6-state regime (period tripling) and the transition between the
two regimes. Figure \ref {iteree_12A} shows the iterated functions
of order 1, 2, 6, and 12. The 6th iterated function crosses the
first diagonal at the same points than the first and the second
iterates only, which means that no 6-state regime is expected. By
contrast, the 12th iterate cuts the diagonal at more points, but
with a very high slope, indicating that the corresponding fixed
points cannot be stable. This, combined with the fact that no
convergence to a square function (constant by domains), such as
$f^{16}$ in Figure \ref {iteree_16}, suggests an aperiodic
behavior; the time dependent signal shown in
Fig.\ref{figtransit_04445} looks indeed chaotic (nevertheless the
flow always remains positive). The periodic/chaotic character of
the signal can be distinguished by examining the time series, but
a complementary method is the computation of an FFT. For the
signal of Fig.\ref{figtransit_04445}, the spectrum is more regular
than the spectrum of a 6-state periodic regime. Nevertheless the
frequencies of the latter (the ``normal'' frequency $f_{2}$ of the
2-state regime with the frequencies $f_{2}/3$ and $2f_{2}/3$)
remains visible in the spectrum of the first one, as it is often
the case for signals corresponding to very close values of the
parameter. A consequence is that these frequencies clearly appear
when listening the sound.

\begin{figure}[h]
\centering \includegraphics[width=10cm]{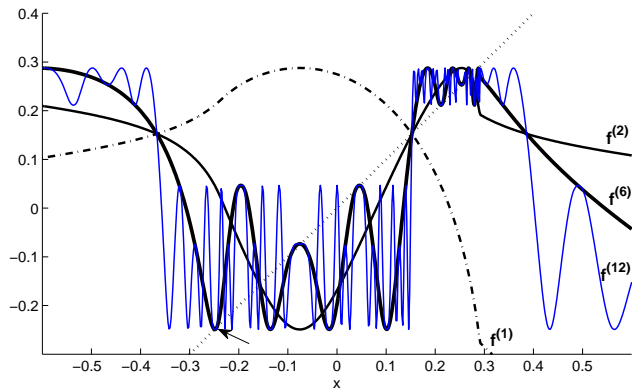}
\caption{{}Iterates for $\protect\lambda =0.95,$ $\ \protect\gamma
=0.4469$ and $\protect\zeta =0.8$, of order 1, 2, 6 and 12. A
convergence to a 6-state regime is observed. The arrow indicates a
region where $f^{(6)}(x)$ cuts the first diagonal.}
\label{iteree_12B}
\end{figure}

\begin{figure}[h]
\centering \includegraphics[width=10cm]{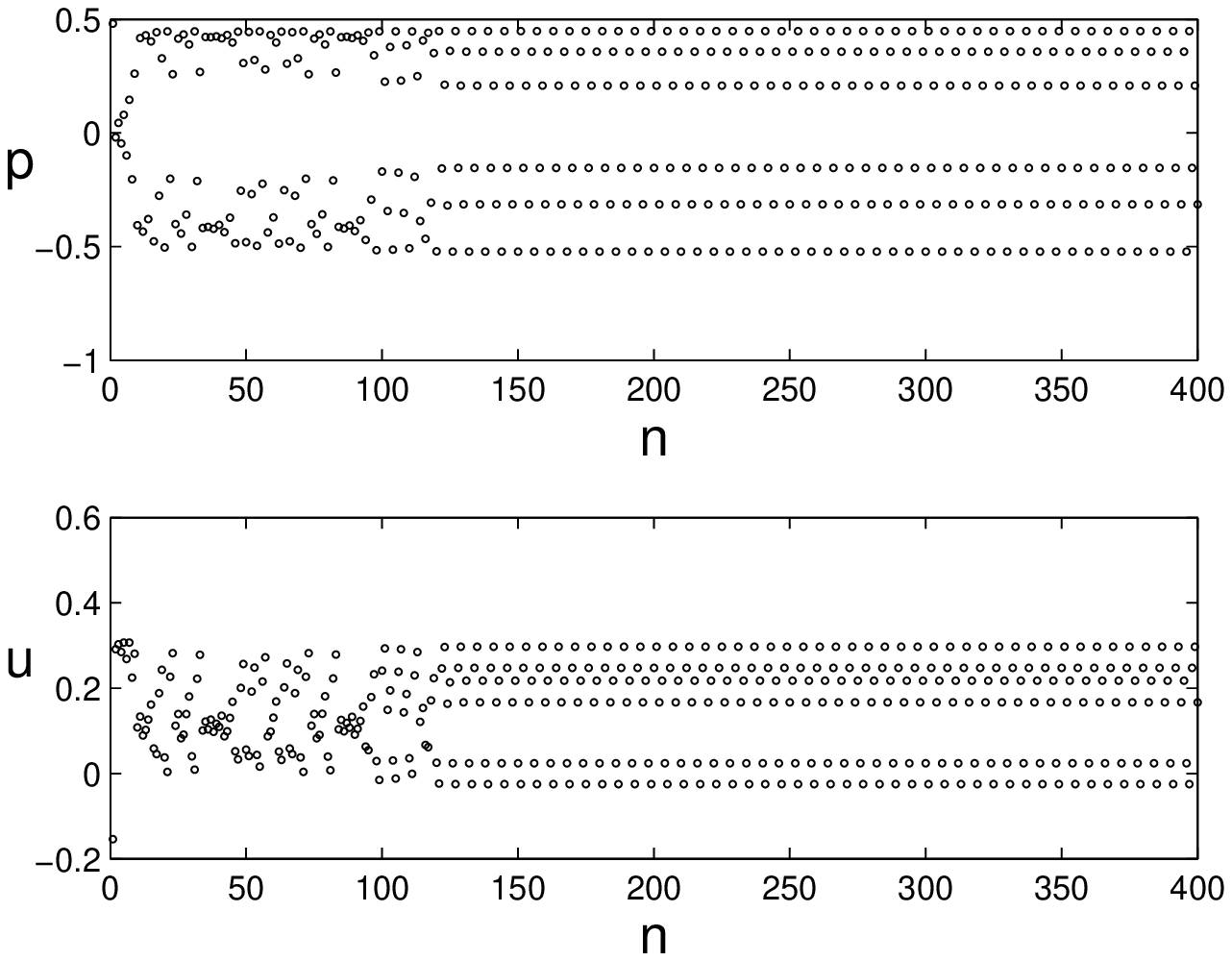}
\caption{{}Iteration from $n=0$ for $\protect\lambda =0.95,$ $\protect\zeta =0.8$, $%
\protect\gamma =0.4469$, $p_{0}^{+}=-0.3347.$ The regime is
periodic (6-state).} \label{figtransit_04469}
\end{figure}

 Figure \ref{iteree_12B} is similar to
figure \ref{iteree_12A}, but with a slightly larger value of
$\gamma $ ($0.4469$ instead of $0.4445$). In the region indicated
by the arrow, one notices that the 6th iterated function now cuts
the first diagonal. They are 12 points of intersection (plus 1
common point with the first iterate as well as two common points
with the second iterate, all unstable); the slope of the tangent
shows that 6 of them are stable, so that one obtains a 6-state,
periodic, regime. The variations of higher order iterates, e.g.
$f^{(12)}$, remain very fast; the convergence to the limit cycle
is then much slower than for Fig. \ref {iteree_16}, except if the
initial point is close to a limit point (e.g. that shown by an
arrow: it turns out that the 12th iterated function is very close
to the 6th one). As a consequence, the initial transient to the
6-state regime can be rather chaotic, as shown in Fig. \ref{figtransit_04469}%
, but convergence to a periodic regime does occur later. This
existence of periodic regimes above the threshold for chaos is
called \textquotedblleft periodicity windows\textquotedblright ,
which appears as a narrow whiter region in Fig. \ref{fig-bif}. A
difference with the usual $2^{n}$-state regimes (when $\gamma $ is
below the chaotic range), for instance corresponding to Fig.
\ref{iteree_gamma=042}, is that one obtains $2^{n}$ intersections
with the diagonal, stable or unstable; by contrast, for the
6-state regime, they are 6 stable and 6 unstable points.

\bigskip

\begin{figure}[h]
\centering \includegraphics[width=10cm]{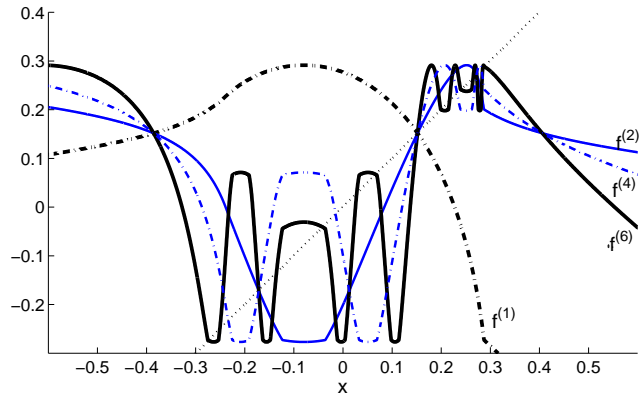}
\caption{{}Iterates for $\protect\lambda =0.95,$ $\ \protect\gamma
=0.4544$ and $\protect\zeta =0.8$, of order 1, 2, 4 and 6. A
convergence to a 6-state regime is observed. } \label{iteree_6A}
\end{figure}

(ii) We now examine the transition between a 6-state regime and a
4-state regime through chaotic regimes or intermittency regimes. For $\gamma =0.4544$%
, a 6-state regime is obtained. Fig. \ref{iteree_6A} shows the iterates of
order 1, 2, 4 and 6. The 4th and 6th iterates have common intersections with
the first and second iterates, since both 4 and 6 are multiples of 2. The
6th iterate intersects the first diagonal at 12 other points, while the 4th
cuts the diagonal at 4 points only. These 4 points are unstable, thus no
4-state regime can exist. On the contrary, for the 6th iterate, half of the
12 points are stable (i.e. with a small slope of the tangent line), so that
one obtains a 6-state stable regime.

\begin{figure}[h]
\centering \includegraphics[width=10cm]{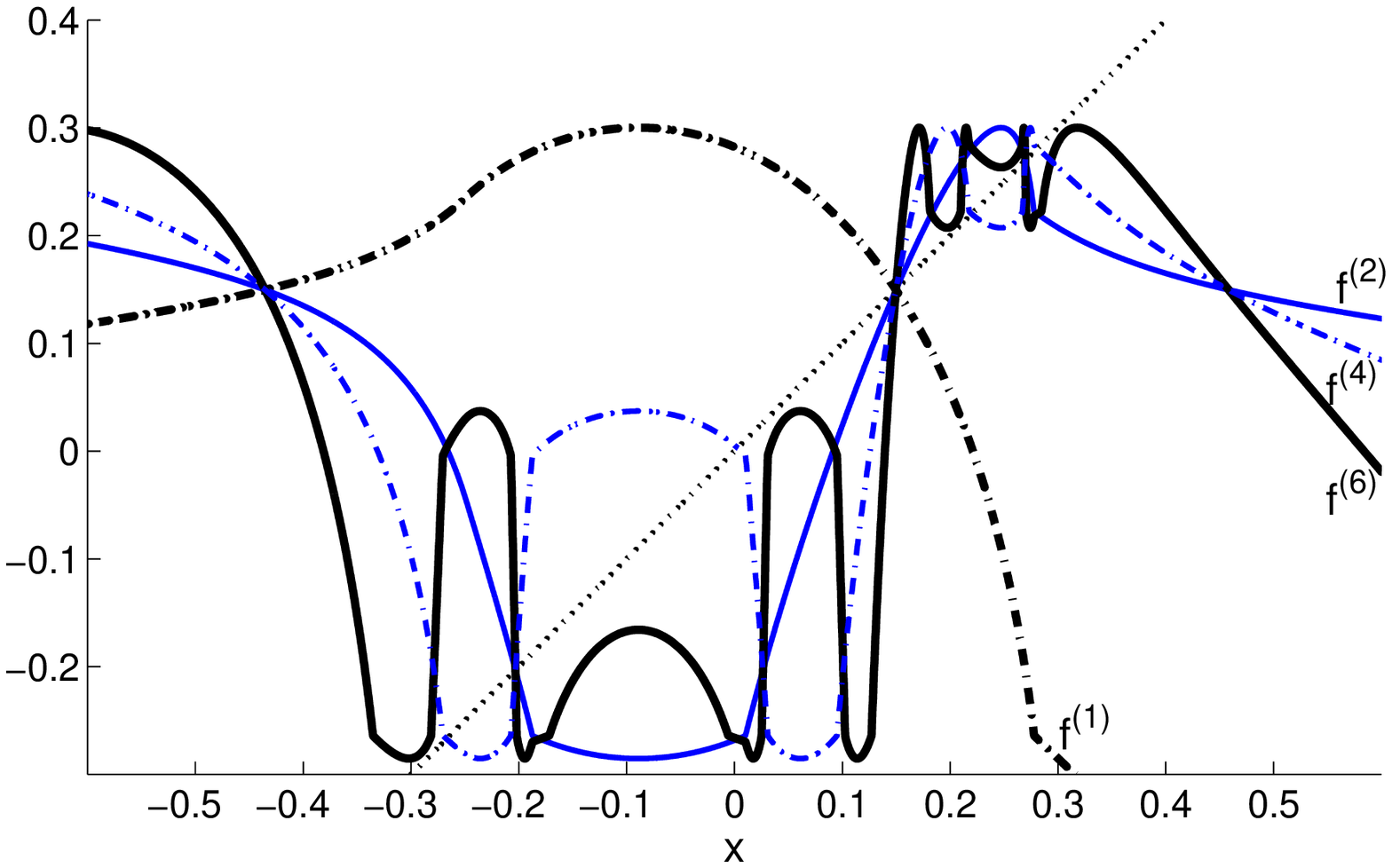}
\caption{{}Iterates for $\protect\lambda =0.95,$ $\ \protect\gamma
=0.472$ and $\protect\zeta =0.8$, of order 1, 2, 4 and 6. A
convergence to a 4-state regime is observed. } \label{iteree_6B}
\end{figure}

\begin{figure}[h]
\centering \includegraphics[width=10cm]{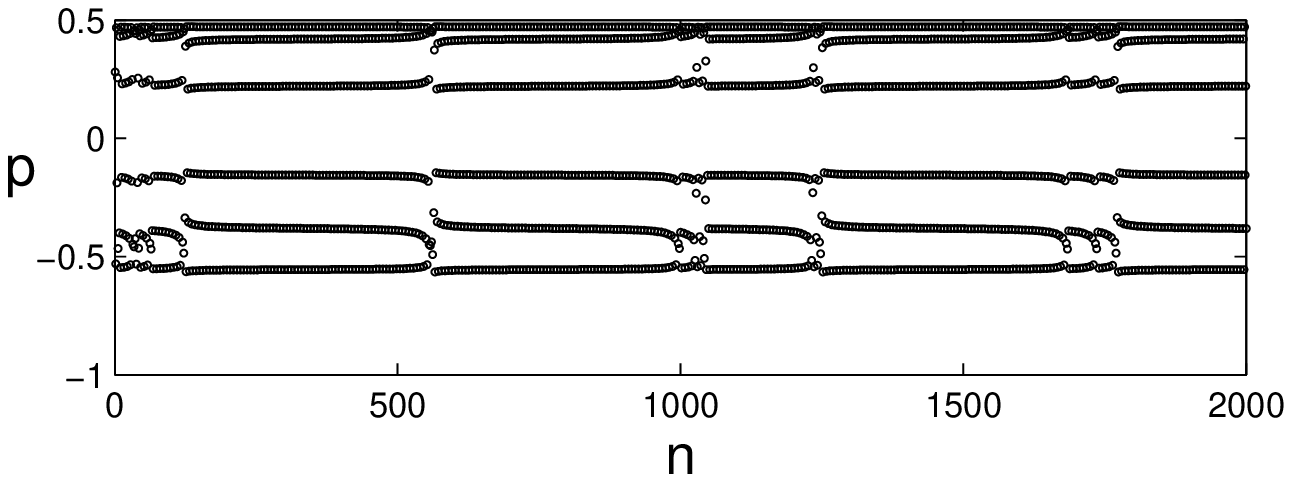} \centering
\includegraphics[width=10cm]{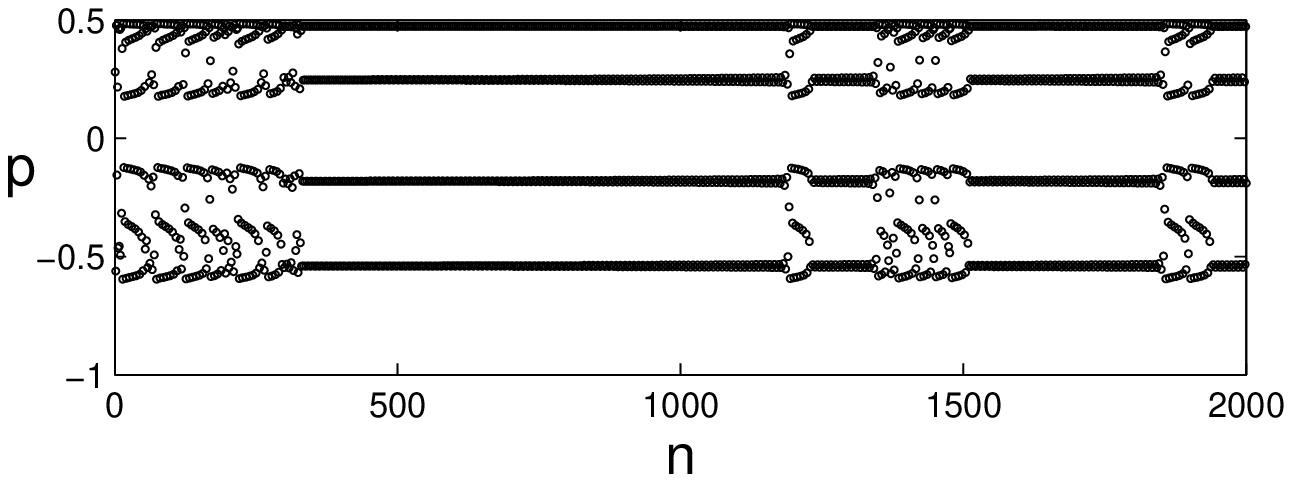}
\caption{{} Iteration from $n=0$ for $\protect\lambda =0.95,$ $\protect\zeta =0.8$, $%
\protect\gamma =0.46623,$ $p_{0}^{+}=0$ (upper curve):
Intermittencies between chaos and a 6-state regime are observed.
However the lower curve (for $\protect\lambda =1,$ $\protect\zeta =0.8$, $%
\protect\gamma =0.467,$ $p_{0}^{+}=0$) shows a more clear
situation of intermittencies between chaos and a 4-state regime. }
\label{figtransit_inter2}
\end{figure}

\begin{figure}[h]
\centering \includegraphics[width=10cm]{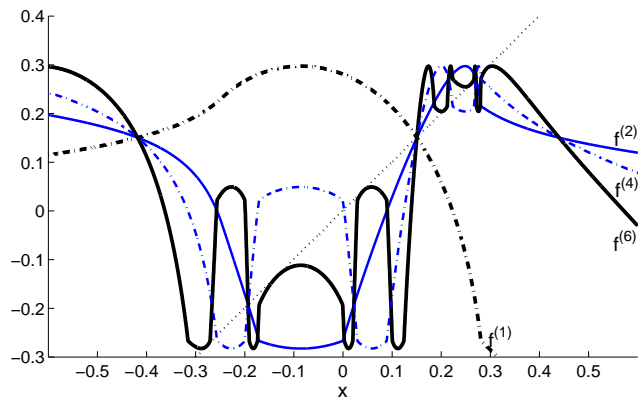}
\caption{{}Iterates for $\protect\lambda =0.95,$ $\ \protect\gamma
=0.46623$ and $\protect\zeta =0.8$, of order 1, 2, 4 and 6,
corresponding to intermittencies. The sixth iterate is tangent to
the diagonal. } \label{iteree_046623}
\end{figure}

What happens for a higher value of $\gamma ,$ namely $0.472$
corresponding to a 4-state regime is shown in Fig.
\ref{iteree_6B}, with again the iterates of order 1, 2, 4, 6. The
4th iterate curve crosses the diagonal for the same number of
points than previously, but the 4 points are now stable. The 6th
order iterate does not intersect the diagonal, except at the
common points with the two first iterates.
 Between the two preceding values
of the parameter $\gamma $, both chaotic and intermittent regimes
can exist. For $\gamma =0.46623$, Figure \ref {figtransit_inter2}
shows intermittencies between a chaotic and a 6-state behaviors
(upper curve), and Figure \ref {iteree_046623} shows that the 6th
iterate is tangent to the first diagonal in 6 points, so that the
resulting permanent regime can be interpreted as a kind of
\textquotedblleft hesitation\textquotedblright\ between two
behaviors. The 4 intersections of the 4th iterate remain
unstable.\\ The lower curve in Figure \ref{figtransit_inter2}
shows another, more visible, example of intermittencies, obtained
with slightly different values of the parameters, between a
chaotic regime and a 4-state one (actually it is a 8-state one,
very close to a 4-state regime).

 \clearpage
\section{Conclusion}

The study of the iteration model of the clarinet should not be
limited to the first iterate: higher order iterates give
interesting information on possible regimes of oscillation. In the
limit of very high orders, their shape gives a direct indication
of the number of states involved in the limit regime, or of
chaotic behavior.\ One can also predict an intermittent regime of
the iterations, which takes place when an iterate is almost
tangent to the first diagonal, so that the iterations are
\textquotedblleft trapped\textquotedblright\ for some time in a
narrow channel. The phenomenon might be related to some kinds of
multiphonic sounds produced by the instrument. It is true that
this phenomenon takes place only in a rather narrow domain of
parameters, but this is also the case of the period doubling
cascade, which has been observed experimentally.\ One can
therefore reasonably hope that the present calculations will be
followed by experimental observations.

\section*{Acknowledgments}

This work was supported by the French National Agency ANR within
the CONSONNES project. We thank also the Conservatoire
neuch\^{a}telois and the high school ARC-Engineering in
Neuch\^{a}tel. Finally we wish to thank Sami Karkar and Christophe
Vergez for fruitful discussions.

\begin{center}
APPENDICES
\end{center}

\appendix{\label{app}}

\section{Analytical iteration function\label{app1}}

\subsection{Derivation of the equations}

Our purpose is to obtain an analytical expression of the iteration function $%
p_{n}^{+}=f(p_{n-1}^{+}).$\ From the basic model (Eqs. (\ref{1} to
\ref{1ter}, \ref{4b-1}, \ref{recurrence})), the following
quantities can be defined:
\begin{eqnarray*}
\text{ }X &=&\gamma -p_{n}=\gamma -p_{n}^{+}-p_{n}^{-}=\gamma
-p_{n}^{+}+\lambda p_{n-1}^{+}\text{\ ; } \\
\text{\ }Y &=&u_{n}+X=\gamma -2p_{n}^{-}=\gamma +2\lambda p_{n-1}^{+}\text{.}
\end{eqnarray*}
$p_{n}^{+}=g($ $p_{n}^{\_})$ can be obtained from the knowledge of the
function $X(Y),$ given by the solving of:
\begin{eqnarray}
&&Y=X\text{ \ (beating reed, }X>1\text{);}  \label{a1} \\
Y &=&X+\zeta (1-X)\sqrt{X}\text{\ (non-beating }  \label{a2} \\
&&\text{reed, positive flow, }0<X<1\text{);}  \notag \\
Y &=&X-\zeta (1-X)\sqrt{-X}\text{ \ \ (non-beating }  \label{a3} \\
&&\text{reed, negative flow, }X<0\text{).}  \notag
\end{eqnarray}
For the non-beating reed case, the study of function $Y(X)$ leads to a
direct analytical solution, as explained below, at least if $\zeta <1$
(otherwise it is a multi-valued function).

Finally, with the notation $x=p_{n-1}^{+}$ and $f(x)=p_{n}^{+}$, if $%
\mathcal{Y}(X)$ is the Heaviside function, the iteration function is
obtained, as:
\begin{eqnarray}
f(x) &=&\gamma -X(Y)+\lambda x\text{, with }Y=\gamma +2\lambda x\text{ and}
\label{a4} \\
Y(X) &=&X+\zeta sign(X)\,\mathcal{Y}(1-X)(1-X)\sqrt{\left\vert X\right\vert }%
.
\end{eqnarray}

\subsection{Non-beating reed, positive flow \ (0$\leq Y\leq 1)$}

For this case, both $X$ and$\ Y$ \ are positive and smaller than unity,
because $\zeta <1.$ Writing $Z=\sqrt{X}$, Eq. (\ref{a2}) is written as:
\begin{equation}
G_{1}(Z)=Y\text{, \ \ where }G_{1}(Z)=-\zeta Z\left[ Z^{2}-\frac{Z}{\zeta }-1%
\right] .
\end{equation}
The study of function $G_{1}(Z)$ shows that it is monotonously increasing
from $0$ to $1$ when $Z$ increases from $0$ to $1.$ Therefore the equation $%
G_{1}(Z)=Y$ has a unique solution when $0\leq Y\leq 1.$ With this condition,
it appears that the equation has three real solutions, and that the
interesting solution (located between $0$ and $1$) is the intermediate one.
As a conclusion, it is possible to use the classical formula for the
solution of the cubic equation:
\begin{eqnarray*}
\sqrt{X} &=&Z=-\frac{2}{3}\eta \sin \left[ \frac{1}{3}\arcsin \left( \frac{%
\psi -\mu }{\zeta \eta ^{3}}\right) \right] +\frac{1}{3\zeta }\text{ \ ;} \\
\psi &=&\frac{1}{\zeta ^{2}}\text{ ; }\eta =\sqrt{3+\psi }\text{ ; }\mu =%
\frac{9}{2}(3Y-1).
\end{eqnarray*}

\subsection{Non-beating reed, negative flow (Y$\leq 0)$}

For this case, both $X$ and$\ Y$ \ are negative$.$ Writing $Z=\sqrt{-X}$,
Eq. (\ref{a3}) is written as follows:
\begin{equation}
G_{2}(Z)=Y\text{, where }G_{2}(Z)=-\zeta Z\left[ Z^{2}+\frac{Z}{\zeta }+1%
\right] .
\end{equation}
The study of the function $G_{2}(Z)$ shows that it is monotonously
decreasing from $0$ when $Z$ increases from $0.$ Therefore the equation $%
G_{2}(Z)=Y$ has a unique real, positive solution when $Y\leq 0.$ The two
other solutions are either real and negative or complex conjugate, with a
negative real part, because the sum of the three solutions is negative ($%
-1/\zeta $). As a conclusion, the solution can be written by using the
following formulae:

\subparagraph{If the discriminant is positive}

\begin{equation*}
discr=q^{3}+r^{2}>0\text{, where}
\end{equation*}
\begin{eqnarray*}
q &=&\frac{1}{9}\left[ 3-\psi \right] \text{ \ ; }r=-\frac{\psi +\mu }{%
27\zeta }\text{ }. \\
\sqrt{-X} &=&Z=s_{1}-\frac{q}{s_{1}}-\frac{1}{3\zeta }\text{; }s_{1}=\left[
r+\sqrt{discr}\right] ^{1/3}.
\end{eqnarray*}

\subparagraph{If the discriminant is negative}

\begin{equation*}
discr=q^{3}+r^{2}<0
\end{equation*}

\begin{eqnarray*}
\sqrt{-X} &=&Z=\frac{2}{3}\eta ^{\prime }\cos \left[ \frac{1}{3}\arccos
\left( -\frac{\psi +\mu }{\zeta \eta ^{\prime 3}}\right) \right] -\frac{1}{%
3\zeta }\text{ ;} \\
\eta ^{\prime } &=&\sqrt{-3+\psi }.
\end{eqnarray*}

\section{Negative flow limit}

\label{app2}

The condition of existence of negative flow is given by $x_{i}>f_{\min }$.\
This is equivalent to the condition on the antecedents, $x_{i}^{\prime
}<f_{\max }$, where $x_{i}^{\prime }$ is the larger antecedent of $x_{i}$,
such as $x_{i}^{\prime }>x_{\max },$ because $f(x)$ is decreasing for all $%
x>x_{\max }$ (see Fig. \ref{fonction-iteree})$.$ Therefore the volume flow
is negative at time $n+1$.

In order to determine the limit value $\gamma _{i}$, the following equations
are to be used:
\begin{eqnarray}
X &=&\gamma -x_{i}+\lambda x_{i}^{\prime }\text{ }=\frac{\gamma }{2\lambda }%
(1+\lambda )^{2}+\lambda A_{\zeta }\text{\ ; } \\
Y &=&\gamma +2\lambda x_{i}^{\prime }=\gamma (1+\lambda )+\lambda 2A_{\zeta
}.
\end{eqnarray}
$\gamma $ being positive (a reasonable hypothesis for the normal playing),
the unknown $X$ needs to be larger than the quantity $\lambda A_{\zeta }.$
Eliminating $\gamma $ in the above equations implies the following equation,
with $X>\lambda A_{\zeta }$:
\begin{equation*}
Y(X)-X=\frac{(\lambda -1)X+2\lambda A_{\zeta }}{1+\lambda },
\end{equation*}
or
\begin{equation}
H(X)=\delta \text{ \ ,}  \label{Eq-H-1}
\end{equation}
with:

\begin{eqnarray}
H(X) &=&(1+\lambda )\left[ Y(X)-X\right] +(1-\lambda )X\text{ \ }
\label{Eq-H-2} \\
\text{for\ \ }X &>&\lambda A_{\zeta }\text{ \ \ ; \ \ }\delta =2\lambda
A_{\zeta }.  \notag
\end{eqnarray}
\begin{figure}[h]
\centering \includegraphics[width=10cm]{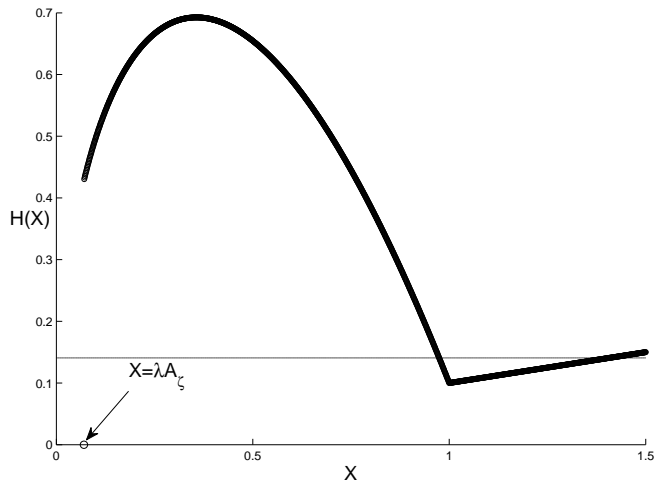}
\caption{Function $H(X)$ given by Eq. (\ref{Eq-H-2}) and constant line $%
\protect\delta =2\protect\lambda A_{\protect\zeta }.$ Two solutions $X>%
\protect\lambda A_{\protect\zeta }$ exits for this case ($\protect\lambda %
=0.9$, $\protect\zeta =0.9$), because condition (\ref{10h}) is satisfied.}
\label{figfonctionH(X)}
\end{figure}
 An example of function $H(X)$ is
shown in Fig. \ref{figfonctionH(X)}. It appears that no solutions
exist if $H(1)>\delta $ and two solutions exist if $H(1)<\delta $,
i.e. if inequation (\ref{10h}) holds. The two solutions can be
obtained analytically.\ However, for sake of simplicity, we give
the exact solution for the larger one, $\gamma _{i2}$, and an
approximation for
the smaller one, $\gamma _{i1}$, obtained at the first order in\ $%
\varepsilon =1-X$:
\begin{eqnarray}
\gamma _{i2} &=&\frac{2\lambda ^{2}A_{\zeta }}{1-\lambda ^{2}}\text{;}
\label{10} \\
\gamma _{i1} &\simeq &\frac{2\lambda }{(1+\lambda )^{2}}\left[ 1-\lambda
A_{\zeta }-\varepsilon \right] \text{,}  \label{10a} \\
&&\text{ with }\varepsilon =\frac{\lambda -1+2\lambda A_{\zeta }}{(\lambda
+1)\zeta +\lambda -1}.
\end{eqnarray}

\bigskip

This error is found to be less than $1\%$ in comparison with the exact
value. Condition (\ref{10h})\bigskip\ can be shown to be necessary and
sufficient. We do not give the entire proof, but it can be shown that
another necessary condition for having two solutions is $H^{\prime
}(1^{-})<0 $, or \ $\zeta (\lambda +1)+\lambda -1>0$, but it is implied by
condition (\ref{10h})\bigskip .

Fig. \ref{limits} shows that the first negative flow threshold $\gamma _{i1}$
is very close to the threshold $\gamma _{b}$, and slightly smaller. For a
given $\lambda $, the limit value of $\zeta $ such as $\lambda
>1/(1+2A_{\zeta })$ corresponds to the equality between the beating reed
threshold and the negative flow one. For a given $\zeta ,$ negative flow is
possible above a certain value of $\lambda $. For rather strong losses, if $%
\lambda <0.84$, no negative flow can occur. For a cylindrical resonator,
this implies that $\alpha \ell >0.085$.



\end{document}